\documentclass[pre,superscriptaddress,tightenlines,twocolumn,showpacs]{revtex4-1}

\usepackage{latexsym,amsmath,amsfonts,tabularx,amssymb,bm,empheq}
\usepackage{subfigure}

\usepackage[normalem]{ ulem }
\usepackage{soul}

\usepackage{color}
\usepackage[dvipsnames]{xcolor}
\definecolor{nblue}{RGB}{28,130,185}

\definecolor{cgreen}{RGB}{76,153,0}

\definecolor{myorange}{RGB}{245,156,74}

\usepackage{hyperref}
\hypersetup{
  colorlinks=true,
  citecolor=magenta,
  urlcolor=-myorange
}

\newcommand{\bea}{\begin{eqnarray}}
\newcommand{\eea}{\end{eqnarray}}

\def\simge{\mathrel{%
   \rlap{\raise 0.511ex \hbox{$>$}}{\lower 0.511ex \hbox{$\sim$}}}}
\def\simle{\mathrel{
   \rlap{\raise 0.511ex \hbox{$<$}}{\lower 0.511ex \hbox{$\sim$}}}}

\def\simle{\mathrel{
   \rlap{\raise 0.511ex \hbox{$<$}}{\lower 0.511ex \hbox{$\sim$}}}}

\def\simge{\mathrel{%
    \rlap{\raise 0.511ex \hbox{$>$}}{\lower 0.511ex \hbox{$\sim$}}}}

\begin{document}

\title{Why Might the Standard Large $N$ Analysis Fail in the O($N$) Model: The Role of Cusps in the Fixed Point Potentials}

\author{Shunsuke Yabunaka}
\affiliation{
Fukui Institute for Fundamental Chemistry, Kyoto University, Kyoto 606-8103, Japan
}

\author{ Bertrand Delamotte}
\affiliation{
Sorbonne Universit\'e, CNRS, Laboratoire de Physique Th\'eorique de la Mati\`ere Condens\'ee, LPTMC, F-75005 Paris, France.
}


\date{\today}

\begin{abstract}
 The large $N$ expansion plays a fundamental role in quantum and statistical field theory. We show on the example of the O$(N)$ model that at $N=\infty$, its  standard implementation misses some fixed points of the renormalization group in all dimensions smaller than four. These new fixed points show singularities  under the form of cusps at $N=\infty$ in their effective potential that become a boundary layer at finite $N$. We show that they have a physical impact on the multicritical physics of the $O(N$) model at finite $N$. We also show that the mechanism at play holds also for the O($N$)$\otimes$O(2) model and is thus probably generic.
\end{abstract}

\maketitle

The $1/N$ expansion is one of the most important tools in field theory. It has played a prominent role in QCD \cite{tHooft1} as well as in statistical mechanics and condensed matter physics \cite{BrezinWallace,Zinn-Justin}. One of its key features is that it can yield reliable results even in strongly coupled models because it is nonperturbative in the coupling constant(s). It also has the enormous advantage of not being linked to a particular dimension, contrary to the usual perturbative expansions. This latter feature has often allowed us to make a bridge between the perturbative expansions performed around the upper and the lower critical dimensions of a model. For instance, at leading order in the $1/N$ expansion both the Mermin-Wagner theorem in two dimensions and the mean-field behavior at criticality in dimensions $d\ge4$ are retrieved, which is out of reach of both perturbative expansions in $\epsilon=4-d$ \cite{4-epsilon} and $\epsilon'=d-2$ \cite{2-epsilon}.





The success of the large $N$ analysis relies on (i) the possibility to extend the original model to arbitrary values of $N$ and (ii) the fact that the model is soluble at $N=\infty$. This is the case not only for the O($N$) and the gauge SU$(N)$ models but also for a large class of  statistical field theories.

We show in this Letter that, surprisingly, even for the O($N)$ model, which is the textbook example for the $1/N$ expansion, the situation is not as simple as it is widely believed. More precisely, we show  on the examples of the O($N$) and O($N)\otimes$O(2) models that at $N=\infty$ some fixed points (FPs) that play an important role even at a qualitative level were missed by the usual large $N$ approach \cite{dattanasio,Tetradis-Litim}. The presence of these fixed points changes  the finite $N$ (multicritical) physics of these models.

Over the years, the importance of renormalization group (RG) FPs  showing cusps has been recognized. This occurs for FP functions such as thermodynamics potentials that are singular for a certain value of their argument. This is the case of the celebrated random field Ising model and is responsible for the breakdown of  supersymmetry and dimensional reduction \cite{tisser08,tisser11}. This is also the case out of equilibrium for  some reaction-diffusion problems \cite{Gredat}. To the best of our knowledge, the occurrence of FPs with a cusp is known only in replica theory applied to disordered systems  and in field theories describing out of equilibrium statistical models (see, however, Ref. \cite{Heilmann}). We prove below that they also play an important role in simple field theories such as the O($N$) and O($N)\otimes$O(2) models since they are responsible for the failure of the usual $1/N$ expansion.

A method of choice for studying the $N=\infty$ limit of the O($N$) model is the computation of the FP effective potentials. This is best achieved by considering Wilson's RG because it is by nature functional. We recall below the takeaway philosophy of the modern version of Wilson's RG known as the nonperturbative  -- or functional -- renormalization group (NPRG).

The NPRG is based on the idea of integrating
fluctuations step by step \cite{PhysRevB.4.3174}. It is implemented on the Gibbs free energy $\Gamma$ \cite{wetterich91,wetterich93b,Ellwanger,Morris94} of a model defined by a Hamiltonian (or Euclidean action) $H$ and a partition function ${\cal Z}$. To this model is associated a one-parameter family of models with Hamiltonians $H_k=H+ \Delta H_k$ and partition functions ${\cal Z}_k$, where  $k$ is a momentum scale. In $H_k$, $\Delta H_k$ is chosen such that only the rapid fluctuations in the original model, those with wave numbers $\vert q\vert > k$, are summed over in the partition
function ${\cal Z}_k$. Thus, the slow modes ($\vert q\vert < k$) need to be decoupled in ${\cal Z }_k$ and this is achieved by giving them a mass of order $k$, that is by taking for $\Delta H_k$ a quadratic (masslike) term, which is nonvanishing only for the slow modes:
\begin{equation}
 {\cal Z}_k[\boldsymbol{J}]= \int D\boldsymbol\varphi_i \exp(-H[\boldsymbol\varphi]-\Delta H_k[\boldsymbol\varphi]+ \boldsymbol{J}\cdot\boldsymbol\varphi)
 \label{partition}
\end{equation}
with $\Delta H_k[\boldsymbol\varphi]=\frac{1}{2}\int_q R_k(q^2) \varphi_i(q)\varphi_i(-q)$, where, for instance, $R_k(q^2)=(k^2-q^2) \theta(k^2-q^2)$ and $\boldsymbol{J}\cdot\boldsymbol\varphi=\int_x J_i(x) \varphi_i(x)$.
 The $k$-dependent  Gibbs free energy $\Gamma_k[\boldsymbol\phi]$
is defined as  the (slightly modified) Legendre transform of $\log  {\cal Z}_k[\boldsymbol{J}]$:
\begin{equation}
\label{legendre}
 \Gamma_k[\boldsymbol\phi]+\log  {\cal Z}_k[\boldsymbol{J}]= \boldsymbol{J}\cdot\boldsymbol\phi-\frac 1 2 \int_q R_k(q^2) \phi_i(q)\phi_i(-q)
 \end{equation}
with $\int_q=\int d^dq/(2\pi)^d$. With the choice of regulator function $R_k$ above, $\Gamma_k[\phi]$ interpolates between the Hamiltonian $H$ when $k$ is of order of the ultraviolet cutoff $\Lambda$ of the theory,  $\Gamma_\Lambda\sim H$,  and the Gibbs free energy $\Gamma$ of the original model when $k=0$,  $\Gamma_{k=0}=\Gamma$. 
The exact RG flow equation of $\Gamma_k$ gives the evolution of $\Gamma_k$ with $k$ between these two limiting cases and reads \cite{wetterich93b}:
\begin{equation}
\label{flow}
\partial_t\Gamma_k[\boldsymbol\phi]=\frac 1 2 {\rm Tr} [\partial_t R_k(q^2) (\Gamma_k^{(2)}[q,-q;\boldsymbol\phi]+R_k(q))^{-1}],
\end{equation}
where $t=\log(k/\Lambda)$, ${\rm Tr}$ stands for an integral over $q$ and a trace over group indices and $\Gamma_k^{(2)}[q,-q;\boldsymbol\phi]$ is
the matrix of the Fourier transforms of $\delta^2\Gamma_k/\delta \phi_i(x)\delta \phi_j(y)$. 

In most cases, Eq.(\ref{flow}) cannot be solved exactly and  approximations are mandatory. The best-known approximation consists in expanding $\Gamma_k$ in powers of  the derivatives of $\phi_i$ and to truncate the expansion at a given finite order\cite{canet03,canet05,kloss14,delamotte04,benitez08,canet04,tissier10,tisser08,canet16,leonard15}. The approximation at lowest order is dubbed the local potential approximation (LPA). For the O($N$) model, it consists in approximating  $\Gamma_k$ by:
\begin{equation}
 \Gamma_k[\boldsymbol\phi] = \int_{x} \left(\frac{1}{2} (\nabla \phi_i)^2 + U_k(\phi)\right),      
   \label{ansatz-order2}
\end{equation}
where, by definition, $ \phi=\sqrt{\phi_i\phi_i}$. Fixed points are found only for dimensionless quantities and thus we define the dimensionless field $\tilde{\phi}$ and potential $\tilde{U}_{k}$
as 
$\tilde{\phi}=v_{d}^{-\frac{1}{2}}k^{\frac{2-d}{2}}\phi$ and
$\tilde{U}_{k}(\tilde{\phi})=v_{d}^{-1}k^{-d}U_{k}\left(\phi\right)$
with 
$v_{d}^{-1}=2^{d-1}d\pi^{d/2}\Gamma(\frac{d}{2})$. The LPA flow of $\tilde{U}_{k}$ reads: 
\begin{equation}
   \begin{split}
 \partial_{t}\tilde U_{t}(\tilde\phi)=-d\,\tilde U_{t}(\tilde\phi)+\frac{1}{2}(d-2)\tilde\phi\, \tilde U_{t}'(\tilde\phi)+\\  
\left(N-1\right)\frac{\tilde\phi}{\tilde\phi+\tilde U_{t}'(\tilde\phi)}+\frac{1}{1+\tilde U_{t}''(\tilde\phi)}.\hspace{0.5cm}
\end{split}
\label{flow-LPA}
\end{equation}
The usual large $N$ limit of the LPA flow \cite{Tetradis-Litim} is obtained by (i) replacing the factor $N-1$ by $N$, (ii) dropping the last term   in Eq.(\ref{flow-LPA}) because it is assumed to be subleading compared to the term proportional to $N$,  (iii) rescaling the field by a factor $\sqrt{N}$ and the potential by a factor $N$: $\bar\phi=\tilde\phi/\sqrt{N}$, $\bar U=\tilde U/N$. As a consequence of these three steps, the explicit dependence in $N$ of the LPA flow of $\bar U(\bar\phi)$ disappears in the large $N$ limit. The crucial point is that the resulting LPA equation on $\bar U$ can be shown to be  {\it exact} in the limit $N\to\infty$ \cite{dattanasio} (see, however, below). Thus, the problem of finding all FPs of the O($N)$ model. $\partial_{t}\bar U_{t}(\bar\phi)=0$, in the limit $N\to\infty$ boils down to solving the LPA FP equation on $\bar U(\bar\phi)$ having dropped the last term in Eq.(\ref{flow-LPA}). This has been done in detail in several papers \cite{Tetradis-Litim,Katis-Tetradis}. The result is the following: In a generic dimension $d<4$ and apart from the Gaussian FP, there is only one FP  which is the usual Wilson-Fisher (WF) FP. The exception to the rule above occurs  with the the Bardeen-Moshe-Bander FPs that play no role here \cite{Bardeen-Moshe-Bander,David,Omid,Mati2017}.

We now show that the procedure described above is too restrictive and eliminates some FPs that are physically relevant. The point is that the last term in Eq.(\ref{flow-LPA}) is negligible compared to the term proportional to $N-1$ only if it reaches a finite limit when $N\to\infty$. We show that  because of singularities this is not necessarily the case and that the last term in Eq.(\ref{flow-LPA}) can also be of order $N$. 

It is convenient for what follows to change variables. Following Ref. \cite{Morris}, we define: $V(\mu)=U(\phi)+(\phi-\Phi)^2/2$ with $\mu=\Phi^2$ and $\phi-\Phi=-2\Phi V'(\mu)$. As above, it is convenient to rescale $\mu$ and $V(\mu)$:   $\bar\mu=\mu/{N}$, $\bar V=V/N$. In terms of these quantities, the FP equation for $\bar V(\bar\mu)$  reads 
\begin{equation}
 0=1-d\,\bar V+(d-2)\bar\mu \bar V'+4\bar\mu{\bar V'}{}^2-2\bar V'-\frac{4}{N}\bar\mu\,\bar{V}''.
\label{flow-LPA-WP-essai}
\end{equation}
This equation has two remarkable features. First, it is much simpler than Eq.(\ref{flow-LPA}) because the nonlinearity comes only from the $(\bar V')^2$ term. Second,  it is the LPA equation obtained from the Wilson-Polchinski version of the NPRG \cite{Wilson, Polchinski, Hasenfratz, Comellas}: $V$ and $U$ are therefore related by the Legendre transform Eq.(\ref{legendre}). Equation (\ref{flow-LPA-WP-essai}) has therefore also been widely studied in the literature. The usual large $N$ analysis performed in this version of the NPRG consists here again in neglecting the last term in  Eq.(\ref{flow-LPA-WP-essai}) because it is suppressed by a factor $1/N$. Under the assumption that this term is indeed negligible in the large $N$ limit, the resulting equation becomes independent of $N$ and an exact (implicit) solution of this equation is known \cite{Kubyshin} (for concreteness, we plot it in the Supplemental Material). However, at large $N$, it is clear in Eq.(\ref{flow-LPA-WP-essai}) that we have to deal with singular perturbation theory  since the small parameter used for the expansion, that is, $1/N$, is in front of the term of highest derivative, that is,  $\bar V''$. In this case, it is well known that in general the term proportional to $\bar V''$ cannot be neglected and that singular solutions can exist at  $1/N\to 0$ \cite{Holmes}. In other words, at finite but large values of $N$, a boundary layer  can exist  for a particular value of the argument $\bar\mu$ that becomes a singularity at $N=\infty$. We now show that this is what indeed occurs.

It is particularly simple to understand in Eq.(\ref{flow-LPA-WP-essai}) why and how at $N=\infty$ a solution exhibiting an isolated singularity for a given $\bar\mu=\bar\mu_0$  can exist. Consider the intervals on the left and on the right of $\bar\mu_0$. On these intervals, $\bar V(\bar\mu)$ is regular by definition. This implies that the last term in Eq.(\ref{flow-LPA-WP-essai}) can safely be neglected at large $N$ on these two intervals. To get a singularity where this term can play a role in Eq. (\ref{flow-LPA-WP-essai}) at $N=\infty$, 
it is necessary that the solutions on the left and on the right of $\bar\mu_0$  match at $\bar\mu_0$ but with two different slopes. It is trivial to build such a solution: Take on the right of $\bar\mu_0$ the Wilson-Fisher solution of Eq.(\ref{flow-LPA-WP-essai}) (without the last term) and on the left $\bar V(\bar\mu)=\bar{\mu}/2$, which is a trivial solution of Eq. (\ref{flow-LPA-WP-essai}). In each dimension the matching point is the intersection of these two curves. For instance, we have numerically found $\bar\mu_0=0.694$ in $d=3.2$ and $\bar\mu_0\to 1/2$ when $d\to 4^-$.
Then, at finite $N$, the boundary layer around $\bar\mu_0$ where the left and right solutions  match smoothly (but abruptly) can be easily computed at leading order in $1/N$ by  (i) introducing another scaled variable $\tilde{\mu}=N (\bar{\mu}-\bar{\mu}_0)$ and (ii) writing down the FP equation  involving only the leading term in $1/N$ around $\tilde{\mu}=0$, having assumed that the derivative of the slope $\bar V'$  with respect to $\tilde{\mu}$ are of order $1$ around $\tilde{\mu}=0$ (see Sec. I of the Supplemental Material for a complete description of this procedure).
We find that the thickness of the boundary layer is of order $1/N$ in terms of the variable $\bar{\mu}$, which implies that $\bar V''$ is of order $N$ within the layer. This is the reason why the last term in Eq. (\ref{flow-LPA-WP-essai}) is not negligible within the layer. This means that the FP solution $\bar{V}(\bar{\mu})$ does not scale uniformly in $\bar{\mu}$  as $1/N$ -- which is assumed in the usual large $N$ analysis -- but inhomogenously depending on whether $\bar{\mu}$ is located inside or outside the boundary layer. It is important to note that due to this singularity, 
the usual argument  about the exactness of the LPA in the limit $N\to\infty$ is not valid anymore. We have therefore studied the stability of the result obtained above by including the next term of the derivative expansion that consists in replacing $(\nabla \phi_i)^2$ by $ Z_k(\phi)(\nabla \phi_i)^2 + Y_k(\phi)(\phi_i\nabla \phi_i)^2$ in Eq.(\ref{ansatz-order2}): all conclusions drawn with the LPA alone are still valid.

\begin{figure}
\includegraphics[scale=0.9]{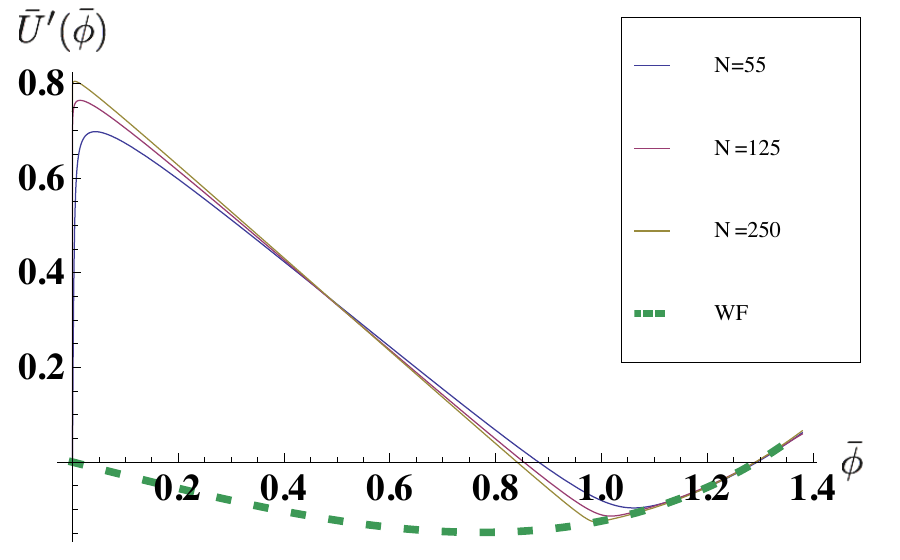}
 \caption{$\bar{U}'(\bar{\phi})$ for the $C_{2}$ FP of Eq. (\ref{flow-LPA}) for different values of $N$  and the Wilson-Fisher FP for $N=100$ in $d=3.2$. }
\label{C2FP}
\end{figure}
Once the boundary layer has been computed from Eq.(\ref{flow-LPA-WP-essai}), it is particularly interesting to transform the FP solution $\bar V(\bar\mu)$ back to $\bar U(\bar\phi)$. For reasons that will be clear in the following, we call $C_2$ this FP. 
 In $d=3.2$, $C_2$ exists and we show it in Fig.\ref{C2FP}. Three interesting features appear on this figure. First, a limiting shape of $C_2$  clearly shows up when $N$ is increased, which is consistent with the existence of a singular FP at $N=\infty$. Second,   for the large values of $\bar\phi$, that is, $\bar\phi>0.965$, $C_2$ coincides with the WF FP whereas it does not at smaller field. Third, between $\bar\phi=0^+$ and  $\bar\phi\simeq0.965$, the slope of $\bar U'(\bar\phi)$ is very close to $-1$, which makes the last term of Eq.(\ref{flow-LPA}) very large. We have checked (i) that this term scales exactly as $N$ at large $N$, and (ii) that, using the relation $\Phi=\phi+\bar{U}'(\bar{\phi})$, the interval $\phi\in[0^+,0.965]$ is exactly mapped onto the (very narrow) boundary layer around $\bar\mu_0$ in the $(\bar\mu, \bar V)$ parametrization. Fourth, at finite $N$,  $\bar U(\bar\phi)$ is a regular function of $\bar\phi^2$ and thus $\bar U'(\bar\phi=0)=0$. Then, $\bar U'(\bar\phi)$ shows an almost vertical slope at large $N$ at $\bar\phi=0$ such that $\bar U'(\bar\phi=0)$ becomes undefined when $N\to\infty$. We have checked using again $\Phi=\phi+\bar{U}'(\bar{\phi})$ that the (very narrow) interval where $\bar U'(\bar\phi)$ varies abruptly around the origin is exactly mapped onto the interval where $\bar V(\bar\mu)=\bar\mu/2$, that is, $[0,\bar\mu_0]$. 
 
A first natural question is to wonder whether  $C_2$ is the only singular FP of the O($N$) model at $N=\infty$. We have found that  $C_2$ appears just below $d=4$ at $N=\infty$ and that, as expected, it does not appear alone but together with another FP that we call $C_3$. The indices 2 and 3 in  $C_2$ and $C_3$ refer to their degree of instability, that is, the number of relevant directions of the RG flow in their neighborhood. These FPs can appear together because their degree of instability differs by one unit. The FP $C_3$ is trivially found from Eq.(\ref{flow-LPA-WP-essai}) at large $N$. It is made of two parts: For $\bar\mu\in[0,2/d]$, $\bar V(\bar\mu)=\bar\mu/2$, and for $\bar\mu\in[2/d,\infty[$, $\bar V(\bar\mu)=1/d$. At finite $N$, these two parts  also connect across a boundary layer of width $1/N$.  When $d\to4^-$, the WF part of $C_2$ at $N=\infty$ corresponding to $\bar{\mu}>\bar{\mu}_0=1/2$ flattens and tends to the value $1/4$. The  potentials of $C_2$ and $C_3$ become identical in this limit which confirms that they coincide in this limit and that they appear together below $d=4$ (see the Supplemental Material where the potentials are plotted).

A second natural question is to wonder whether the FPs found above are nothing but a curiosity occurring at $N=\infty$ with no impact on the physics at finite $N$, in much the same way as the Bardeen-Moshe-Bander FP. We have checked that this is not at all the case. The FPs $C_2$ and $C_3$ found above at $N=\infty$ are indeed the limits of  FPs  found at finite $N$ \cite{Yabunaka-Delamotte-PRL2017}.  These FPs  are regular for all values of the field. They play a prominent role for the multicritical physics of the O($N$) model at least for sufficiently large values of $N$. In particular, their presence solves a paradox: It is well known that the perturbative tricritical FP $T_2$, found perturbatively for all $N$ in $d=3-\epsilon$, is not found at $N=\infty$ for $d<3$. This paradox is solved when realizing that $T_2$ appears for any $N$ at $d=3^-$  where it is Gaussian, and, when $N$ is large enough,  disappears  when decreasing $d$ by colliding with $C_3$ on a line $d_c(N)$ in the $(N,d)$ plane whose equation is $d_c(N)\simeq 3- 3.6/N$ \cite{Yabunaka-Delamotte-PRL2017,Pisarski, Osborn}. Thus, the interval in $d$ where it exists shrinks to 0 when $N$ increases. We notice that both $C_2$ and $C_3$ exist at finite and large $N$ in $d=3$ and it would be very interesting to find models whose multicritical behavior is described by these FPs.

 
A third natural question is whether what we have found is specific to the O($N$) model or is likely to be generic. We have pragmatically investigated the O($N)\otimes$ O(2) model along the same line as above to answer this question.

The order parameter of the O($N)\otimes\,$O(2) model is the $N\times2$ matrix $\Phi=\left(\mathbf{\boldsymbol{\varphi}}_{1},\mathbf{\boldsymbol{\varphi}}_{2}\right)$ \cite{Delamotte-review, Kawamura-review,Yosefin} and the  Hamiltonian is the sum of the usual kinetic terms and of the potential $U(\rho,\tau)$ where $\rho$ and $\tau$ are the two O$(N)\otimes\,$O(2) independent invariants: $\rho=\mathbf{\boldsymbol{\varphi}}_{1}^2+\mathbf{\boldsymbol{\varphi}}_{2}^2$ and $\tau=((\mathbf{\boldsymbol{\varphi}}_{1}^2-\mathbf{\boldsymbol{\varphi}}_{2}^2)^2/4+ (\mathbf{\boldsymbol{\varphi}}_{1}\cdot\mathbf{\boldsymbol{\varphi}}_{2})^2)$. The LPA ansatz is identical to Eq. (\ref{ansatz-order2}) up to the replacement $U_k(\phi)$ by $U_k(\rho,\tau)$. The standard large $N$ limit predicts that, aside from the $O(2N)$-symmetric FP, two nontrivial FPs exist in $2<d<4$: the chiral fixed point $C_+$, which describes the second order transition between the ordered and the disordered phases, and the antichiral fixed point $C_-$, which is tricritical \cite{Jones,Bailin,Gracey,Pelissetto}. Since the LPA equation for the potential is much more involved than in the O$(N)$ model, we have decided to expand the FP potential $U(\rho,\tau)$ around its  minimum $\tilde{\kappa}$:
\begin{equation}
\tilde{U}\left(\tilde\rho,\tilde\tau\right)=\sum_{n,m}\frac{1}{m!n!}\tilde{a}_{mn}\left(\tilde{\rho}-\tilde{\kappa}\right)^{m}\tilde{\tau}^{n},
\end{equation}
where $\tilde{a}_{mn}$ are  coupling constants. The O($2N$) FP is retrieved by setting $\tilde{a}_{mn}=0$ for $n\ge1$ and by rescaling the couplings according to
\begin{equation}
\tilde{\kappa}  =  N^{-1}\bar\kappa,\ \ \ 
\tilde{a}_{mn}  =  N^{-m-2n+1}\bar{a}_{mn},
\label{eq:newscaling2-2}
\end{equation}
which is a direct consequence of the usual rescaling: $\bar{\mathbf{\boldsymbol{\varphi}}}_{i}=\tilde{\mathbf{\boldsymbol{\varphi}}}_{i}/\sqrt{N}$ and $\bar U=\tilde U/N$. Once $\bar{U}$ is Taylor expanded around its minimum, the exactness of the LPA flow at  $N=\infty$ translates in a hierarchical structure of the flows of the couplings: for instance, for the O($2N$) invariant flow, the flow of $\bar{a}_{m0}$ depends only on the set of couplings $\{\bar{a}_{p0}\}$ with $p\le m$. The system of FP equations of the $\bar{a}_{m0}$'s is therefore close and soluble whatever the value of $m$ and the couplings $\{\bar{a}_{m,0}\}$ are said to be  ``perfect coordinates" \cite{Aoki}. The same holds true for  $C_+$ with the rescaling (\ref{eq:newscaling2-2}): at $N=\infty$, the FP equations of the couplings $\bar{a}_{mn}$ with $2m+4n\le2l$ depend only on couplings $\bar{a}_{m'n'}$ with $2m'+4n'\le2l$. 

For $C_-$, the situation is different because   the couplings do not satisfy Eq. (\ref{eq:newscaling2-2}) with finite $\bar{a}_{mn}$'s in the limit $N\to\infty$. By studying numerically the behavior of these couplings, we have found  the proper scaling for the $C_-$ couplings:
\begin{equation}
\tilde{a}_{0n}  =  N^{-2n+1}\bar{a}_{0n},\ 
\tilde{a}_{mn}  =  N^{-m-2n}\bar{a}_{mn}\ (m\ne0).
\label{eq:eq:newscaling3}
\end{equation}
As a consequence of these scalings, the $\bar{a}_{mn}$'s are not  perfect coordinates for $C_-$ and the LPA is therefore not  exact when $N\to\infty$. However, by using an ansatz including all kinds of second and fourth order derivative terms, we have checked in detail that the scaling Eq. (\ref{eq:eq:newscaling3}) is valid independently of the LPA and remains the same beyond this approximation (See Sec. III of Supplemental Material for the details of this ansatz \cite{Supp}). We note that if were using for $C_-$ the usual rescaling Eq. (\ref{eq:newscaling2-2}): $\bar{\mathbf{\boldsymbol{\varphi}}}_{i}=\tilde{\mathbf{\boldsymbol{\varphi}}}_{i}/\sqrt{N}$ and  $\bar U=\tilde U/N$, we would find that the scaled couplings $\bar{a}_{mn}\ (m\ne0)$ vanish in the limit $N\to\infty$. This would invalidate the large $N$ analysis based on this scaling.

The scaling Eq. (\ref{eq:eq:newscaling3}) together with the fact that the $\bar{a}_{mn}$'s are not  perfect coordinates at large $N$  for $C_-$ has dramatic consequences that we now describe. When $d$ is decreased towards $d=3$, we numerically find that the FP couplings $\bar{a}_{mn}$ at $C_-$ diverge. For instance, $\bar{a}_{20}$, and $\bar{a}_{11}$ diverge as $(d-3)^{-1/2}$ while $\bar{a}_{12}$,  $\bar{a}_{21}$, and $\bar{a}_{30}$ as $(d-3)^{-1}$. We have also computed the four most relevant  eigenvalues $\sigma_{1,\cdots,4}$ of the flow at $C_-$, and we have found at $N=\infty$,
$\sigma_{1}=-2,\,\sigma_{2}=d-4,\,\sigma_{3}=2(d-3),\,\sigma_{4}=4-d$.
when $d\to 3^+$ \cite{note1}. In our convention, a negative eigenvalue corresponds to a relevant direction. We conclude that when $d\to3^+$, the first irrelevant eigenvalue vanishes while the coordinates of the FP diverge. This clearly suggests that (i) $C_-$ collides with another FP in $d=3$ when $N=\infty$ and then  disappears below $d=3$ and that (ii) the coordinates of the other FP do not scale with $N$ as in (\ref{eq:eq:newscaling3}), which explains that the collision can occur only if the coordinates $\bar{a}_{mn}$ of $C_-$  no longer have a finite limit when $d\to3^+$. We have looked for other  rescalings than Eq. (\ref{eq:eq:newscaling3}) yielding other FPs and we have found two such FPs that we call $M_2$ and $M_3$, whose coordinates scale as: 
\begin{equation}
\tilde{a}_{0n} =  N^{-1}\bar{a}_{0n}\, ,\, 
\tilde{a}_{mn}  =  N^{-\frac{m}{2}-\frac{3n}{2}}\bar{a}_{mn} \ \ (mn)\neq(01).
\label{eq:eq:newscaling3-1}
\end{equation}
We have checked that $M_2$ and $M_3$ appear simultaneously at $N=\infty$ below  $d\simeq3.37$ and that $C_-$ and $M_3$ collide in $d=3$ and both disappear below this dimension. Neither $M_2$ nor $M_3$ are perturbative FPs since they are never infinitesimally close to the Gaussian FP. We have checked that as in the O($N$) model these two FPs exist at finite $N$ and are physically relevant. They are indeed responsible for the disappearance of  $C_-$ on a line $d_c(N)$, which, in turn, explains why $C_-$ is not found around $d=2$ in the $\epsilon'=d-2$ expansion of the O($N)\otimes\,$O(2) nonlinear sigma model \cite{Azaria}. We notice that within the standard $1/N$ analysis, not only $M_2$ and $M_3$ are not found but $C_-$ is not found to disappear below $d=3$ for $N=\infty$. We conclude that, contrary to the O($N$) model, the standard large $N$ analysis does not only miss some FPs but predicts the existence of $C_-$ for $2<d<3$ where, in fact, it does not exist. It is intriguing to notice that at $N=\infty$ and in both models it is exactly in $d=3$ that the perturbative tricritical FP, either $T_2$ or $C_-$, disappears.

To conclude, we have found new FPs in both the O($N$) and O($N)\otimes\,$O(2) models   whose effective potentials  show singularities  at $N=\infty$ and boundary layers at finite $N$. This makes it difficult to find them with the usual toolbox of the $1/N$ approach. We have also shown that these FPs play an important role in the multicritical physics of these models. In particular, some of them collide with the standard, that is, perturbative tricritical FPs in some dimension, which makes these latter FPs disappear. It is still an open question to have an exhaustive classification of all possible singular FPs of the O($N$) model at $N=\infty$ and in particular of the subset of these FPs that survive and play a physical role at finite $N$. We conjecture that what we have found for the tricritical FPs repeats for all the multicritical FPs. It is also an open and intriguing question to understand why the new FPs found above all have at least two directions of instability and whether new FPs with only one unstable direction could exist. It would also be interesting to integrate the flow to obtain the phase diagram in the presence of all these FPs.

S. Y. was supported by Grant-in-Aid for Young Scientists (B) (15K17737 and 18K13516). We acknowledge  H. Chat\'e, S. Sasa, N. Tetradis and N. Wschebor for discussions and/or  advice about the manuscript. 

\newpage

\widetext

\begin{center}
\textbf{\large Supplemental Materials\\
Why can the standard large $N$ analysis fail in the O($N$) model: The role of cusps in the fixed point potentials} 
\end{center}

\section{Matching procedure and boundary layer analysis for $C_2$}

\begin{figure}[ht]
\includegraphics[scale=0.7]{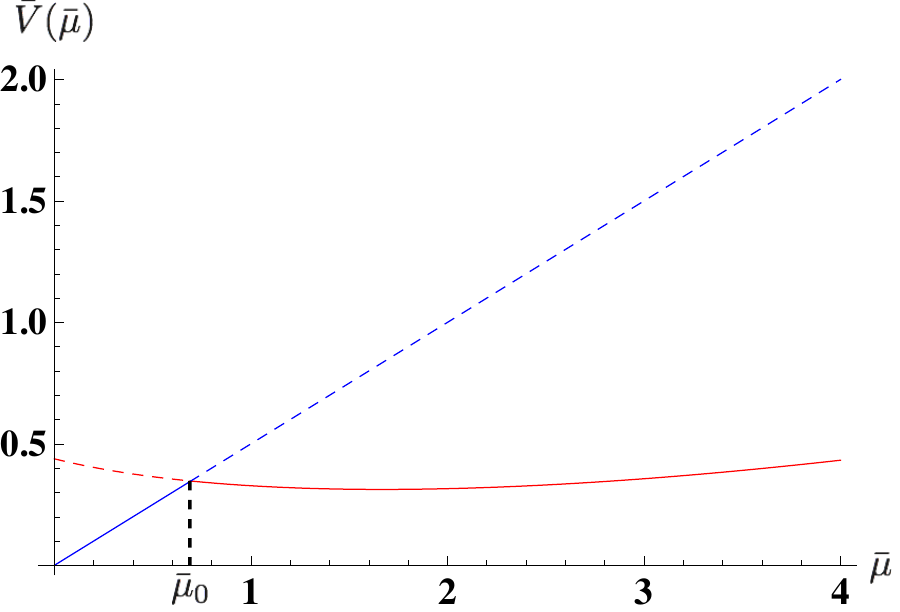}
\caption{ The FP $C_2$ of Eq. (6) in $d=3.2$ at $N=\infty$. It is shown  as a solid line and is made of two parts that match at $\bar{\mu}_0=0.694$. The  part on the right of $\bar{\mu}_0$ is identical to the WF FP.}
\label{matching}
\end{figure}

\begin{figure}[h]
\includegraphics[scale=0.7]{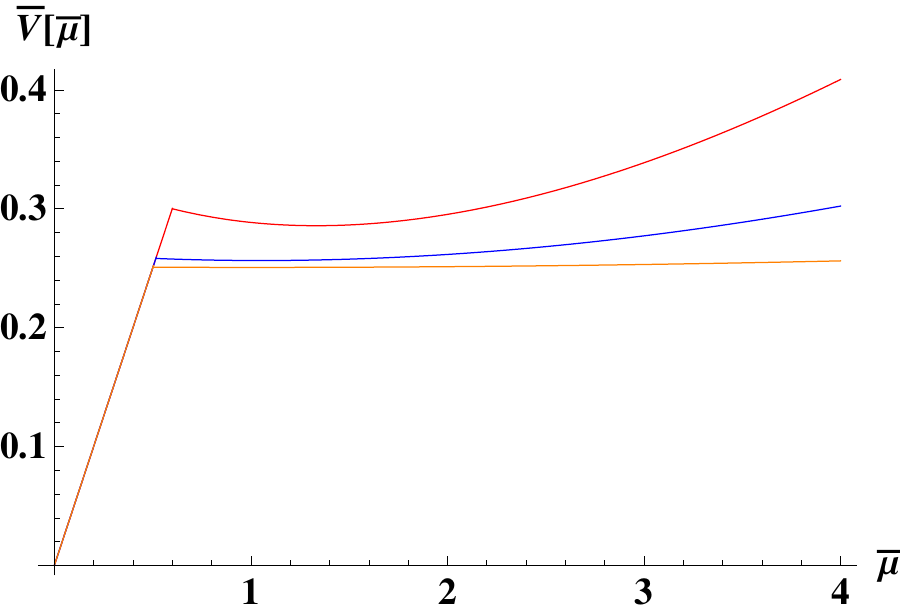}
\includegraphics[scale=0.53]{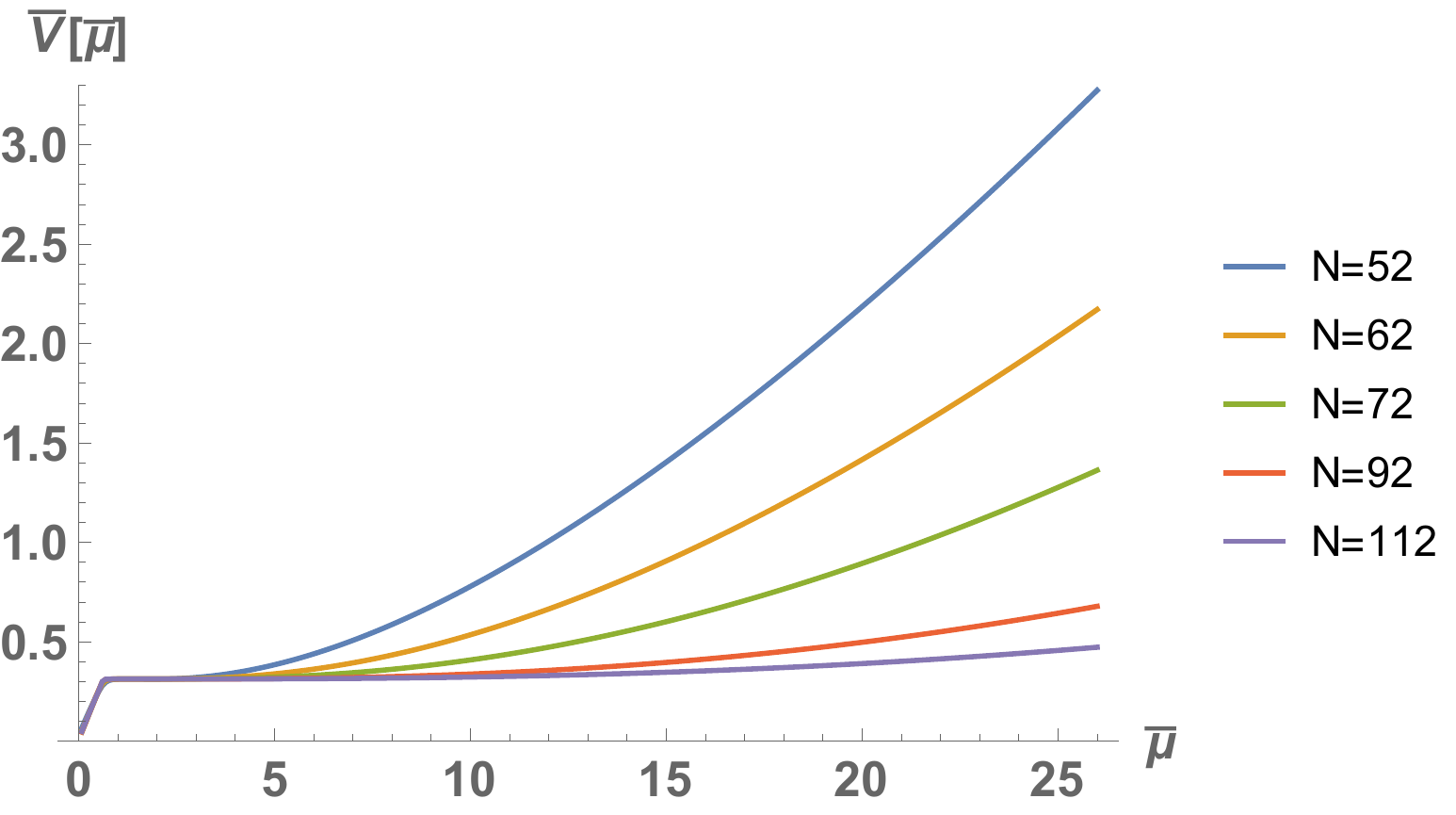}
\caption{Left: The $C_2$ FP  of Eq. (6) in $d=3.5$ (red), $d=3.9$ (blue) and $d=3.99$ (orange) at $N=\infty$. For $\bar{\mu}>\bar{\mu}_0$, it coincides with the  WF FP solution and thus becomes flat when  $d\to4^-$. Right: The $C_3$ FP  of Eq. (6) in $d=3.2$ for various values of $N$. It becomes flat for $\bar\mu>2/d$ in the large $N$ limit. At $N=\infty$, the location of the cusp in both potentials goes to 1/2 when $d\to4^-$ and the two FPs coincide.}
\label{2solutions}
\end{figure}
We recall that the standard Large-$N$ limit of Wilson-Polchinski's version of the FP equation in LPA is given by
\begin{equation}
 0=1-d\,\bar V+(d-2)\bar\mu \bar V'+4\bar\mu{\bar V'}{}^2-2\bar V'.
\label{flow-LPA-WP-essai}
\end{equation}
The exact (implicit) solution of Eq. (\ref{flow-LPA-WP-essai}) which corresponds to the  Wilson-Fisher (WF) FP at $N=\infty$ is known [35]  and plotted in Fig. \ref{matching} in $d=3.2$ along with the trivial solution $\bar{V}(\bar{\mu})=\bar{\mu}/2$. As explained in the letter, we can identify $\bar\mu_0$ as the intersection of the two curves as shown in Fig. \ref{matching}. 




Let us now describe the detail of the boundary layer analysis of Eq. (6). We assume that $\bar{V}'(\bar{\mu})$ remains of order 1 and changes its value from $\bar{V}'(\bar{\mu}_0^-)=1/2$, for the trivial solution $\bar{V}(\bar{\mu})=\bar{\mu}/2$, to $\bar{V}'(\bar{\mu}_0^+)=-0.0794$, for the WF FP solution. This occurs  across the thin boundary layer located at $\bar\mu_0$, whose width is of order  $1/N$ so that $\bar {V}''$ scales as $N$. Inside this boundary layer, we introduce a scaled coordinate $\tilde{\mu}=N (\bar{\mu}-\bar{\mu}_0)$ and denote the slope $\bar{V}'(\bar{\mu})$ by $F(\tilde\mu)$. Then, starting from Eq. (6), we can write down a differential equation which is  valid inside this boundary layer  at the leading order in $1/N$ as  
\begin{align}
   \begin{split}
0=1-d\,\bar V(\bar{\mu}_0)+(d-2)\bar\mu_0 F +4\bar\mu_0{F}^2-2 F-4\bar\mu_0\,F',
\end{split}
\label{flow-LPA-WP}
\end{align}
where $\bar\mu$ in $-d\,\bar V(\bar{\mu})$, $\bar\mu F$, $\bar\mu F^2$ and $\bar\mu F'$  has been replaced by $\bar{\mu}_0$. The primes in Eq. (\ref{flow-LPA-WP}) stand for derivatives with respect to the scaled variable $\tilde{\mu}$. This differential equation has a solution, 
\begin{eqnarray}
F(\tilde{\mu})=V_1 - V_2 \tanh(V_2 \tilde{\mu}),
\end{eqnarray}
where we have defined  $V_1=1/4+\bar{V}'(\bar{\mu}_0^+)/2$ and $V_2=1/4-\bar{V}'(\bar{\mu}_0^+)/2$. This boundary layer solution smoothly (but abruptly) connects the two values $\bar{V}'(\bar{\mu}_0^-)=1/2$ and $\bar{V}'(\bar{\mu}_0^+)=-0.0794$ across the boundary layer, as expected.

\section{Shape  of $C_2$ and $C_3$ at $N=\infty$ in the Wilson-Polchinski parametrization}

We show in Fig. \ref{2solutions} the two FP potentials of $C_2$ and $C_3$ that are solutions at $N=\infty$ of Eq. (6). At $N=\infty$ and for $C_2$, $\bar{\mu}_0\to1/2$ when $d\to4^-$ and for $\bar{\mu}>\bar{\mu}_0$ it flattens and $\bar V(\bar\mu_0)\to 1/4$. As for $C_3$, the potential is flat at $N=\infty$ for any $d>3$ for $\bar\mu>2/d$ as can be seen in Fig. \ref{2solutions}.  At $N=\infty$, the location of the cusp in both potential goes to 1/2 when $d\to4^-$ and the two FPs coincide. Above $d=4$ neither of these FPs exist.

\section{Effects of the higher order derivative terms for O$(N)\otimes\,$O(2) models}

\label{sec:derivateves}

In order to see whether our results about $C_-$, $M_2$ and $M_3$ are modified by higher order derivative
terms, we considered the following ansatz for the effective action. It  includes the  second and fourth order derivative terms that are leading in the large $N$ limit and that therefore could play a role.
\begin{eqnarray}
\Gamma_{k}\left[\mathbf{\boldsymbol{\varphi}}_{i}\right] & = & \int d^{d}\mathbf{x}\left(\frac{1}{2}\left(Z_{0k}+Z_{1k}\left(\rho-\kappa\right)\right)\left[\left(\partial\boldsymbol{\varphi}_{1}\right)^{2}+\left(\partial\boldsymbol{\varphi}_{2}\right)^{2}\right]+\frac{1}{2}Z_{2k}\left[\left(\partial^{2}\boldsymbol{\varphi}_{1}\right)^{2}+\left(\partial^{2}\boldsymbol{\varphi}_{2}\right)^{2}\right]\right.\nonumber \\
 &  & +\frac{\omega_{k}}{4}\left(\boldsymbol{\varphi}_{1}\cdot\partial\boldsymbol{\varphi}_{2}-\boldsymbol{\varphi}_{2}\cdot\partial\boldsymbol{\varphi}_{1}\right)^{2}+\frac{1}{4}Y_{k}^{\left(2\right)}\left(\boldsymbol{\varphi}_{1}\cdot\partial\boldsymbol{\varphi}_{1}+\boldsymbol{\varphi}_{2}\cdot\partial\boldsymbol{\varphi}_{2}\right)^{2}\nonumber \\
 &  & \left.+\frac{1}{4}Y_{k}^{\left(3\right)}\left(\left(\boldsymbol{\varphi}_{1}\cdot\partial\boldsymbol{\varphi}_{1}-\boldsymbol{\varphi}_{2}\cdot\partial\boldsymbol{\varphi}_{2}\right)^{2}+\left(\boldsymbol{\varphi}_{1}\cdot\partial\boldsymbol{\varphi}_{2}+\boldsymbol{\varphi}_{2}\cdot\partial\boldsymbol{\varphi}_{1}\right)^{2}\right)+U_{k}\left(\rho,\tau\right)\right).
\end{eqnarray}
$Z_{0k},Z_{1k},Z_{2k},\omega_{k},Y_{k}^{\left(2\right)}$ and $Y_{k}^{\left(3\right)}$
are coupling constants that do not depend on $\mathbf{\boldsymbol{\varphi}}_{i}$
$(i=1,2)$. For LPA, $Z_{1k},Z_{2k},\omega_{k},Y_{k}^{\left(2\right)}$ and $Y_{k}^{\left(3\right)}$ are set to $0$. Using this ansatz, we solved the FP equation and found
that these additional terms do not almost change
the LPA results on $d_c(N)$ for $N\gtrsim20$, which strongly suggests that LPA results are
very precise in the limit $N\rightarrow\infty$.


\begin{thebibliography}{1}
\bibitem{note1} To show this result analytically, it is convenient to expand the potential around the origin not the minimum.
\end{thebibliography}
\end{document}